\documentclass[openacc]{rstransa}

\usepackage{xcolor,cancel}

\definecolor{g-blue}{rgb}{0.83,0.95,1}
\definecolor{g-yellow}{rgb}{1,1,0.7}
\definecolor{g-green}{rgb}{0.9,1,0.9}
\definecolor{green}{rgb}{0,0.6,0}
\definecolor{cyan}{rgb}{0,0.7,0.7}
\definecolor{black}{rgb}{0,0,0}
\definecolor{grey}{rgb}{0.4 ,0.4 ,0.4 }

\def\blue#1{\textcolor{blue}{#1}}

\usepackage{bm}
\usepackage{amssymb}
\usepackage{amsfonts}
\usepackage{amsmath}
 \usepackage{graphicx}
  \usepackage{amsmath,bm,epsfig}
\def \ed {\end{document}}
\def\Fbox#1{\vskip1ex\hbox to 8.5cm{\hfil\fboxsep0.3cm\fbox{%
  \parbox{8.0cm}{#1}}\hfil}\vskip1ex\noindent}  

\newcommand{\eq}[1]{(\ref{#1})}
\newcommand{\Eq}[1]{Eq.\,(\ref{#1})}
\newcommand{\Eqs}[1]{Eqs.\,(\ref{#1})}
\newcommand{\Fig}[1]{Fig.\,\ref{#1}}
\newcommand{\Figs}[1]{Figs.\,\ref{#1}}
\newcommand{\Sec}[1]{Sec.\,\ref{#1}}
\newcommand{\Refn}[1]{Ref.\,\cite{#1}}

\def\be{\begin{equation}}
\def\ee{\end{equation}}
\def\bea{\begin{eqnarray}}
\def\eea{\end{eqnarray}}
\def\bse{\begin{subequations}}
\def\ese{\end{subequations}}
\newcommand{\BE}[1] {\begin{equation}\label{#1}}

\let \= \equiv  \let\*\cdot \let\~\widetilde \let\^\widehat \let\-\overline
\let\p\partial

  \def\1{\bm1} 

\def\<{\left\langle}    \def\>{\right\rangle}
\def\({\left(}          \def\){\right)}
 \def \[ {\left [} \def \] {\right ]}

\renewcommand{\a}{\alpha}

\newcommand{\ve}{\varepsilon}

\newcommand{\B}[1]{{\bm{#1}}}
\newcommand{\C}[1]{{\mathcal{#1}}}    

\renewcommand{\sb}[1]{_{\text {#1}}}  
\renewcommand{\sp}[1]{^{\text {#1}}}  
\def\Sb#1{_{\scriptscriptstyle\rm{#1}}}
\def\He4 {$^4$He~}

\begin{document}

\title{Theory of anisotropic superfluid $^4$He counterflow turbulence
 }
\author{  Victor S.  L'vov$^{1}$,  Yuri V. Lvov $^{2}$,  Sergey Nazarenko$^{3}$, and Anna Pomyalov$^{1}$}
\address{$^{1}$Dept. of Chemical and Biological Physics, Weizmann Institute of Science, Rehovot, Israel\\
$^{2}$Rensselaer Polytechnic Institute, Troy NY 12180 USA\\
$^{3}$Institut de Physique de Nice, Universit\'e Cote d'Azur, CNRS, Nice, France}

\subject{Statistical hydrodynamics, He-4 superfluids}


\corres{Anna Pomyalov\\
\email{anna.pomyalov@weizmann.ac.il}}

\begin{abstract} 
 We develop an analytic theory of strong anisotropy of the energy spectra  in the thermally-driven  turbulent counterflow of superfluid $^4$He. The key ingredients of the theory are the three-dimensional differential closure for  the vector of the energy flux and the anisotropy of the mutual friction force.  We suggest an approximate analytic solution of the  resulting  energy-rate equation, which is fully supported by the numerical solution.  The  two-dimensional energy spectrum is strongly confined in the direction of the  counterflow velocity. In agreement with the experiment, the energy spectra in the direction orthogonal to the counterflow exhibit two scaling ranges: a near-classical non-universal cascade \\ dominated range and a universal critical regime at large wave-numbers. The theory predicts the dependence of various details of the spectra and the transition to the universal critical regime  on the flow parameters.   
 
 This article is part of the theme issue ‘Scaling the turbulence edifice'.
\end{abstract}

\maketitle

\section*{\label{s:intro}Introduction}

 Most universal properties of turbulence are only revealed in flows with very high Reynolds number. Typically, such conditions are found in atmospheric turbulence or in very large wind tunnels. 
 Liquid helium has very low kinematic viscosity and, therefore, becomes
an ideal test-bed for high-Reynolds-number turbulence even in a relatively small experimental facility.
 The liquid helium viscosity decreases with temperature and 
 below  the Bose-Einstein condensation temperature $T_\lambda\approx 2.17\,$K  $^4$He becomes superfluid. In this state, it can be described as a two-component fluid in which a viscous normal-fluid  and an inviscid superfluid components interact via a mutual friction force\cite{Donnely,2,Feynman,Tough,Vinen,37,SS-2012,TenChapters, BLR,NemirReview}.  

 Various ways of turbulence generation in superfluid He produce flows with very different properties. Mechanically-driven superfluid He with two components flowing in the same direction and coupled by the mutual friction  almost at all scales, is long considered similar\cite{SS-2012,TenChapters,BLR} to the classical flows\cite{Fri}. The similarity included the behaviour of the structure functions and scaling of the turbulent energy spectra close to $k^{-5/3}$\cite{Tab,Roche-new,DNS-He4,Salort2012,Gorgio2016,BLR, Varga-2018,WG-2021}.  

 The two-fluid nature of the superfluid $^4$He  allows generation of turbulence by thermal gradient. In such a flow, that has no classical analogy, the two fluid components flow in opposite directions: the normal fluid carries the heat flux away from the heat source, while the superfluid flows towards the heater\cite{2,Tough,Vinen,TenChapters,NemirReview,Vinen2,Vinen3}. The mutual friction force that couples the components, leads for both the energy exchange and additional dissipation by mutual friction  that are scale-dependent\cite{decoupling,LP-2018}. Since all relevant fluid parameters\cite{DB98} are strongly temperature-dependent, the statistical properties of such a counterflow are not universal.
 Instead, the statistics of the counterflow depends
 on the temperature and on the relative velocity $\B{U}\sb{ns}$\cite{Coexistance,WG-2015,WG-2017,LP-2018,He4-PRL-DNS,CF-DNS}.   
 Recent flow visualization experiments\cite{WG-2015,WG-2017, WG-2018, Prague1,Prague2} stimulated theoretical and numerical investigations of the energy spectra of the counterflow turbulence. It was shown\cite{decoupling,LP-2018,CF-DNS,TsubotaGuo-2020, Giorgio-2020} that besides the dependence on  flow parameters, the energy spectra are sensitive to the angle with respect to the direction of the counterflow velocity. As a result, the energy spectra in the counterflow turbulence are anisotropic and strongly suppressed in the direction of $\B U\sb{ns}$.  

 Although such a spectral anisotropy was predicted theoretically and  confirmed numerically\cite{CF-DNS,He4-PRL-DNS}, the experimental investigations of the energy spectra  for the time being are limited to the plane, orthogonal to the direction of the counterflow velocity\cite{WG-2017,WG-2018}, while the theory  of counterflow turbulence\cite{LP-2018} was developed assuming spectral isotropy. In this paper we relax this assumption and offer a theoretical description  of the spectral anisotropy of the energy spectra of the counterflow turbulence in superfluid $^4$He. 
 
The paper is organized as follows.
 In the  \Sec{s:theory} we develop the theory of anisotropic turbulence. 
Similar to our previous studies  of superfluid turbulence\,\cite{He4,DNS-He4,LP-2018,DNS-He3,decoupling}, we describe the large-scale turbulence in superfluid $^4$He   by  the coarse-grained  Navier-Stokes Equations\,\eqref{NSE} coupled by the mutual friction force. These equations are detailed in 
\Sec{s:theory}(a).
In
\Sec{s:theory}(b)
we introduce some statistical characteristics of anisotropic turbulence, used in our paper. In the focal 
\Sec{s:theory}(c),
we  suggest  the  energy rate \Eqs{final}  for the  axially-symmetric counterflow turbulence.
The key element [Sec.1(c,ii)] 
in the resulting energy rate \Eqs{final} is the   cross-correlation  function $D(k_\|)$  , which  depends only on $k_\|$, according to \Eqs{LP-20}.
In Sec.1(c,iii)
we introduce a vector energy flux $\B \ve(\B k)=\{\ve_\|(\B k), \ve_\perp(\B k)\}$, which depends now on the position in the plane $\B k=\{k_\|,k_\perp\}$, formed by the components  $k_\|$ and $k_\perp$ of the wavevector $\B k$, parallel and orthogonal to the counterflow velocity $\B U\sb{ns}$, respectively.  
We analyze the resulting energy rate equation analytically in \Sec{ss:isotrop2D} and numerically in \Sec{s:sol}.
Finally, in \Sec{s:sum} we summarize our findings.

 \section{\label{s:theory} A theory of anisotropic counterflow turbulence } 
The superfluid phase of liquid  He is characterized by quantized vorticity that is constrained to vortex-line singularities of core radius $a_0\approx  10^{-8}\,$cm and  fixed circulation $\kappa= h/M$, where $h$ is Planck's constant  and $M$ is the mass of the \He4 atom\cite{Feynman}. The  superfluid turbulence is manifested  as a complex tangle of these vortex lines   with a typical inter-vortex distance\cite{Vinen}   $\ell\sim 10^{-4}- 10^{-2}\,$cm. 

 Large-scale hydrodynamics of such a system is usually described by a  two-fluid model, interpreting $^4$He as a mixture of two coupled fluid components: an inviscid  superfluid and a viscous normal fluid.  The temperature-dependent densities of the components $\rho\sb s, \rho\sb n: \rho\sb s+ \rho\sb n=\rho$ define their contributions to the mixture. Here $\rho$ is the density of $^4$He. The fluid components  are coupled by the mutual friction force, mediated by the tangle of quantum vortexes\,\cite{Donnely,Vinen,Vinen2,BLR,Vinen3,37}. 
 
\subsection{\label{ss:HVBK}	Coarse-grained equations for   counterflow	 He-4 turbulence}

Similar to \cite{CF-DNS}, our approach to the problems of large-scale counterflow turbulence  \cite{He4,decoupling,LP-2018}  is based on the 
	coarse-grained  equations\cite{He4,DNS-He4,LP-2018,DNS-He3}  of the incompressible superfluid turbulence. 	These equations   have a form of two Navier-Stokes equations (NSE) for the turbulent velocity fluctuations of the normal fluid and superfluid components $ 
u\sb n(\B r,t)$ and $\B u\sb s(\B r,t)$  in the presence of space-homogeneous mean normal and superfluid velocities  $ 
\B U\sb n $ and $\B U\sb s $:
\begin{subequations}\label{NSE} \begin{align}   \label{NSEs} 
  \frac{\p \,\B u\sb s}{\p t}+  [(\B u\sb s+\B U\sb s)\*
\B\nabla] \B u\sb s
- \frac 1{\rho\sb s }\B \nabla p\sb s  &  =\nu\sb s\,  \Delta \B u\sb s   + \B f
\sb {ns} \,, \quad \B f\sb {ns}  \simeq \Omega\sb s  \,(\B  u \sb n-\B  u \sb s ) \,, 
\\  \label{NSEn}
  \frac{\p \,\B u\sb n}{\p t}+[(\B u\sb n+\B U\sb n) \* \B
\nabla]\B u\sb n
- \frac 1{\rho\sb n }\B \nabla p\sb n & = \nu\sb n\,  \Delta \B u\sb n
-\frac{\rho\sb s}{\rho\sb n}\B f \sb {ns} \, ,   \quad \Omega\sb s    =  \a (T)  \kappa \C L \,,\end{align}\end{subequations} 
 coupled by the mutual friction force $\B f\sb{ns}$ in the
form \eqref{NSEs}.  It also involves  the
temperature dependent dimensionless dissipative mutual friction parameter $\alpha(T)$ and the
superfluid   vorticity  $\kappa \C L$. Here $\C L$ is the vortex line density
(VLD). Furthermore, the partial densities of the components are $\rho\sb s,\rho\sb n$, $ p\sb n   =\frac{\rho\sb n}{\rho }[p+\frac{\rho\sb s}2|\B u\sb
s-\B u\sb  n|^2]\,, \quad
p\sb s =  \frac{\rho\sb s}{\rho }[p-\frac{\rho\sb n}2|\B u\sb s-\B u\sb
n|^2] $ denote the pressure of the normal and the
superfluid components,  the kinematic
viscosity  of normal fluid component $\nu\sb n=\eta / \rho \sb n$ with $\eta$
being the dynamical viscosity\cite{DB98}  of normal \He4 component  and the Vinen's effective superfluid viscosity\,\cite{Vinen}  $\nu\sb s$,
which  accounts\cite{He4} for the energy dissipation at the intervortex
scale $\ell$ due to vortex reconnections, the energy transfer to  Kelvin waves and other dissipation mechanisms

We consider here the planar heat source, typically used in the channel counterflow.
 
\subsection{\label{ss:stat}Statistical characteristics of anisotropic turbulence } 

The general  description of the homogeneous superfluid \He4 turbulence at the level of the second-order statistics can be done in terms of the three-dimensional (3D) correlation functions of the normal-fluid and superfluid turbulent velocity fluctuations in the $\B k$-representation:   
 \begin{align} \begin{split} \label{def-Fb} 
(2\pi)^3\delta^3(\B k -\B k') \C F^{\alpha\beta}   _{ij} (\bm k) = \<    v _i^\alpha (\bm k)\cdot      v_j^{*\beta} (\bm k')\>\,,   
   \quad \C F_{ij}(\bm k) \=  \sum_{\alpha=x,y,z}  \C F^{\alpha\alpha}_{ij}(\bm k)   \ . 
\end{split}  \end{align}     
Here $\C F_{j}^{\alpha\beta}(\bm k)=   \C F_{jj}^{\alpha\beta}(\bm k)$,    $\delta^3(\B k -\B k')$ is 3D Dirac's  delta function and   
\begin{align} \label{def-Fc} 
\B v_j(\B k)&= \int \B u _j(\B r)\, \exp(-i\B k\cdot \B r)\, d\B r\,,\quad 
\B u_j(\B r) = \int \B v _j(\B k)\, \exp(-i\B k\cdot \B r)\,  d\B k/(2\pi)^3\ .
 \end{align}
 The subscripts ``$_{i,j}$"  denote  the normal
(${i,j}=$n)  or  the superfluid (${i,j}=$s) fluid components and  $^*$ stands for complex conjugation.    The 3D correlation function $\C F   _{ij} (\bm k)$ and the Fourier transform\,\eqref{def-Fc} are  defined such that the kinetic energy density per unit mass $ E_j$ (with the dimension $[E]=$cm$^2$/s$^2$) reads 
$$  E_{j }= \frac12   \< |\B u _j(\B r)|^2 \>=   \frac 12  \int   \C F _{jj}  (\bm k) \frac{d^3 k}{(2\pi)^3}$$.

Due to the presence of the preferred direction, defined by the counterflow velocity $\B U\sb{ns}$, the  counterflow turbulence has an axial symmetry  around that  direction. In this case,   $\C F_{ij}^{\alpha\beta}(\B k)$ \  depends only on two     projections $k_\|$ and $k_\perp$  of the wave-vector $\B k$:    $\B k_\|\= \B U\sb{ns} (\B k\cdot \B U\sb{ns})/ U\sb{ns}^2$ and $\B k_\perp=   (\B k - \B k_\|)$, being independent of the angle $\varphi$ in the $\perp$-plane, orthogonal to $\B U\sb{ns}$: $\C E_{ij}^{\alpha\beta}(\B k)\Rightarrow \C E_{ij}^{\alpha\beta} (k_\|,k_\perp)$.    

In the case of axial symmetry,   a two-dimensional (2D) object $ E^{\alpha\beta}  _{ij} (k_\|,k_\perp)$  still contain all the information about 2$\sp{nd}$-order statistics of the counterflow turbulence:  
$  \displaystyle  E   _{j} (k_\|,k_\perp) \=    \frac{k_\perp}{4\pi^2} \C F   _{j} (k_\|,k_\perp)$.
Now the total kinetic energy  density per unit mass  can be found as
$  \displaystyle E_j=  \int \hskip -.2 cm \int   \limits _0^\infty d  k_\| \,  d    k_\perp E_j ( k_\|,  k_\perp  )$.
 In fully isotropic case, $E_j ( k_\|,  k_\perp  )$ depends only on $k=\sqrt{k_\|^2+k_\perp^2}$ and we can introduce traditional one-dimensional (1D) energy spectrum
 \begin{equation}\label{1D}
 \~ E_j(k)= 2\pi k E_j ( k_\|,  k_\perp  )\ . 
 \end{equation} 

\subsection{\label{ss:theoryC}Energy rate equations for counterflow turbulence}
\subsubsection{General form of the energy rate equation in  axial symmetry }
A theory of space-homogeneous counterflow turbulence\,\cite{LP-2018}, developed under simplifying assumption of  the spectral isotropy of the flow, is based on the stationary balance equations for the 1D energy spectra $\~ E_j(k)$, \Eq{1D}.  Here, we relax  the assumption of the isotropy,  and  derive an energy rate   equation for the 2D energy spectra  $E_j( k_\|,  k_\perp  )$ of the counterflow turbulence with axial symmetry around $\B k_\|$. 
To this end,  we, following\,\cite{LP-2018},  eliminate the pressure terms  using the incompressibility conditions,   Fourier transform  and multiply them by the complex conjugates of the corresponding velocities. After ensemble averaging,  we  get the equations for the 3D spectra  $\C F_j(\B k)$, defined by \Eq{def-Fb}, and average them only over the azimuth  angle $\varphi$  in the plane orthogonal to  $\B k_\|$.  Finally, we get: 
 \begin{align} \label{balance1} 
\frac{\partial E_j(\B k,t)}{\partial t}+\mbox{div}_{\B k} [\B \ve_j ( \B k)]   =&  \Omega_j \big
[E\sb{ns}( \B k  ) - E _j (\B k ) \big ]   - 2\, \nu_j k^2 E_j(\B k)\,, \quad \Omega\sb n = \frac{\Omega\sb s \rho\sb s}{ \rho \sb n} \ . 
  \end{align} 
Here $\B k=\{ k_\|,k_\perp\}$ is a 2D wavevector, $\B \ve_j ( \B k)=\{ \ve_j^{\|}, ( \B k),\ve_j^{\perp} ( \B k)\}$ is the vector of the energy flux.
The cross-correlation function $E\sb{ns}$ is discussed in the next section and the derivation of the vector energy flux is detailed in \Sec{sss:transfer}.
\subsubsection{\label{sss:decor} Cross-correlation function in counterflow turbulence}
 
In our analysis, we use the model of the anisotropic cross-correlation function $E\sb{ns}(k_\|,k_\perp)$,  introduced by Eq.(13) of \Refn{decoupling}:
\begin{subequations}\label{cross}	\begin{equation}\label{LP18c}
	E\sb{ns}(\B k)= \frac{A(\B k)\Omega\sb{ns}}{\Omega\sb{ns}^2+ ( k_\|  U\sb{ns})^2}\,, \quad A(\B k)= \Omega\sb s E\sb n (\B k)+ \Omega\sb n E\sb s (\B k)\,, \quad \Omega\sb{ns}=\Omega\sb n + \Omega \sb s\ .
	\end{equation}
Further simplifications \cite{LP-2018}, allow one to rewrite \Eq{LP18c} for $E\sb{ns}(\B k)$ in the following form: 
 \begin{align} \label{LP-20}
	E\sb{ns}(\B k)&=  E_j(\B k)\big [1-D(k_\|)\big]\,, \quad 
	  D(k_\|)  = {k_\times^2}/\big (k_\times^2+k_\|^2\big ) \,, \quad k_\times= {\Omega\sb{ns}}/{U\sb{ns}}   \ .
 	\end{align}\end{subequations}
 	 Note, that while substituting   $E\sb{ns}(\B k)$ into the rate \Eq{balance1},  we should take in \Eq{LP-20} $j=$n in the equation for the normal component, and $j=$s for the superfluid component.

The physical meaning of the two-dimensional decorrelation function $D(k_\|)$ in \Eq{LP-20} is the same as in the spherical case: it describes the level of decorrelation of the  normal-fluid and superfluid velocity components by the counterflow velocity.  For $k_\|\lesssim  k_\times$, normal-fluid and superfluid velocities are almost fully coupled. In this case, the mutual friction only weakly affects the energy balance. The energy spectrum in the inertial interval of scales is determined by the step-by-step cascade energy transfer. Accordingly, this range of wavenumbers can be called ``cascade-dominated"\cite{LP-2018}.   For large $k_\|$,   $D(k_\|) \ll  1$ and  the velocities of fluid components  are almost decoupled.  In this  ``mutual-friction dominated range", the energy dissipation by mutual friction  strongly suppresses the energy spectra.   

\subsubsection{\label{sss:transfer} The energy transfer term}

The energy transfer term div$_{\B k} [\B \ve_j ( \B k)]$  in \Eq{balance1} originates from the
nonlinear terms in the coupled NSE \Eqs{NSE}  and has the same form\cite{LP-95,LP-2,BL} as in the classical
turbulence:
 \label{Tr}
	\begin{align}  \begin{split}
	\label{genA}
  \mbox{div}_{\B k}  [\B \ve_j(\B k)]  &\= 	\frac{d \B \ve_j(\B k)}{d \B k} =2\, \mbox{Re}\Big\{\int V^{\xi\beta\gamma}(\B k,\B q,\B
	p)   \C E_j^{\xi\beta\gamma} (\B k,\B q,\B p)\, \delta(\B k+\B q+\B p)\frac{d^3 q \, d^3 p}{(2\pi)^6} \, \Big\}\,, 
	\\ 
  V^{\xi\beta\gamma} (\B k,\B q,\B p) & =  i  \Big ( \delta _{\xi \xi'}- \frac{ k^\xi
		k^{\xi'}}{k^2}   \Big ) 
 \Big( k^\beta \delta_{\xi ' \gamma} + k^\gamma \delta _{\xi' \beta}
	\Big )\ .
	\end{split}	\end{align}
 Here $\C E_j^{\xi\beta\gamma}(\B k,\B q,\B p)$ is the  simultaneous
triple-correlation function of turbulent (normal or superfluid) velocity fluctuations
in the $\B k$-representation, that  we will not specify here and $
V^{\xi\beta\gamma}(\B k,\B q,\B p)$ is the interaction vertex in the NSE.
Importantly,  the right-hand-side of \Eq{genA} conserves the total
turbulent kinetic energy (i.e. the integral of $E_j(k)$ over entire $\B k$-space)
and therefore can be written in the divergent form  as div$_{\B k} [\B \ve_j ( \B k)]$.

A simple algebraic closure approximation for the  energy flux $\~\ve(k)$ in  isotropic turbulence follows from the dimensional reasoning in the framework of Kolmogorov 1941 (K41) hypothesis\,\cite{Fri}: 
 \begin{subequations}\label{K41}
   \begin{equation}\label{K41A}
 \~\ve(k)= \~C k^{5/2} \~ E^{3/2} (k) \ .
 \end{equation}
Here $\~ C$  is a dimensionless  constant of the order of unity and $\~\ve$ is the  energy flux  in the inertial interval of scales.  The equation \eq{K41A} immediately gives the celebrated  $\frac 53$-law:
$\displaystyle \~ E\Sb{K41}(k)= C\Sb{K41} \~ \ve^{2/3} k^{-5/3}$
with $C\Sb{K41}= \~C ^{-2/3}$. The experimental value~\cite{C-K41} of the constant $C\Sb{K41}\simeq 0.5$.
In the 2D case with axial symmetry along the counterflow direction,  the situation is  more involved. Now, the 2D vector $\B \ve$ with the dimensions  $[\ve]=$(cm/s)$^3$  is the flux of 2D-energy density $E(\B k)$ per unit mass per square of unit $k$ with the dimensions $[E]$=cm$^4$/s$^2$. The dimensional reasoning, similar to that leading to \Eq{K41A}  gives 
 \begin{equation}\label{K41B}
 |\ve(\B k)| \approx C  k^3 E_j^{3/2} (\B k )\, , \quad \B k=\{ k_\|,  k_\perp  \}\,,
 \end{equation}
 \end{subequations}
 with $C=\~C /\sqrt{2\pi}\simeq 1.1$.
 
 Unfortunately, the dimensional reasoning  does not allow us to reconstruct the direction of the vector $\B \ve$. It is natural to assume that $\B \ve$  is oriented in the direction of the   steepest descent of the 3D energy spectrum, 
 i.e.  along $ \nabla_{\B k} \big[E(\B k)/k\big]$ or, if this gradient is zero,  $\ve =0$.  Note that this allows to   satisfy an  additional physical requirement that the energy flux vanishes in the  thermodynamic equilibrium with equipartition of energy, when $E(k)\propto k$\cite{Leith}. Thus, requiring the Kolmogorov-type scaling properties, we choose the energy flux in the form  $\ve \propto \B \nabla_{\B k} \big[E(\B k)/k\big]^{3/2}$.   Reconstructing the prefactor according to \Eq{K41B}, one finds 
 \begin{subequations}\label{closure}
 \begin{equation} \label{K41D}
\B \ve (\B k) = - C_1\, k^{11/2}   \B \nabla_{\B k} \Big[ \frac{E (\B k)}k \Big]^{3/2}\,,  \quad \B \nabla_{\B k}\= \frac{d}{d\B k}\,, 
 \end{equation} 
 with some new dimensionless coefficient $C_1\approx 2\, C/11\simeq 0.2$. The numerical factor is chosen such that closures \eqref{K41B} and \eqref{K41D} coincide for K41 spectrum $E(k)\propto k^{-8/3}$. In the isotropic 2D case, $\ve(k)\propto 1/k$. This gives $E(k)\propto k^{-8/3}$, as required.  
  
 It was shown previously\cite{WG-2015,WG-2017,WG-2018,LP-2018,He4-PRL-DNS,CF-DNS} that the energy spectra in the counterflow do not have a simple power-law form in the inertial interval.
 To account for that it was proposed\cite{LP-2018} to replace $C_1$ by  a  function $C_1(\B k)$ that depends on the local slope of the spectrum. Here, we use the same approach and introduce the coefficient
 \begin{equation}  \label{K41E}  
 C_1(\B k)=\frac{4~ C_1}{3[4-m(\B k)]}\, ,\quad m(\B k)=- \B k \cdot \B \nabla_{\B k} \ln E(\B k)\, ,
 \end{equation}
 \end{subequations}
    that depends self-consistently  on the local slope $ m(\B k)$  of the energy spectra in the steepest descent direction.  The function $\B C_1(\B k)$ increases when $m$ approaches the critical value $m=4$, at which the energy transfer over scales looses its locality and, formally, $\ve\to \infty$.  
  
 For $m>4$,  the  energy flux in a range from some $\tilde  k$ to $k\gg \tilde k$ becomes non-local (similar to $^3$He) and  requires a more sophisticated closure\, \cite{DNS-He3}.

 \subsubsection{\label{sum} Final form of the energy rate equation }
Combining  \Eqs{balance1} with \Eqs{LP-20}, \eqref{K41D}, and   \eqref{K41E} and  neglecting the viscosity term,  in the stationary case we finally have
 \begin{align} \label{final} 
 \cancel{\frac{\partial E_j(\B k,t)}{\partial t}} \blue{-}    \B \nabla_{\B k }\cdot  \Big \{C_{1j}(\B k)\,k^{11/2}   \B \nabla_{\B k } \Big[ \frac{E_j (\B k)}k \Big]^{3/2}\Big \}  =- \frac{\Omega_j E_j(\B k)   \,k_\|  ^2}{ k_\| ^2+ k_\times^2}\,,  \quad \B k=\{k_\|,k_\perp \}\,. 
   \end{align} 

 Recall that $\Omega\sb s=\alpha \kappa \C L$, $\Omega\sb n= \Omega\sb s \rho\sb s/\rho\sb n$, and  $\Omega\sb {ns}= \Omega\sb s \,\rho/\rho\sb n$. The crossed term with time derivative is preserved here (and in some equations below) to stress that this is a continuity equation for the energy spectrum. In theoretical analysis we will use only stationary version of this (and similar) equations, while numerically we consider its full version and look  for its stationary solutions by numerically integrating continuity equation 
 from appropriate initial conditions. 
 
 To simplify the appearance 
of  the energy rate \Eqs{final}  and to open a way to its numerical solution, we introduce a new function $\Psi_j(\B q,t)$ instead of 
$E_j(\B k,t)$:
\begin{equation}\label{Psi1}
    E(\B k)= E(k_0)\,\Psi^2(\B q)\, q^{-8/3}\,,\quad \B q\= \B k/k_0\,,
\end{equation}
such that  the fast K41 dependence of $  E(\B k)$ is explicitly accounted for:  with K41 scaling $\Psi(\B q)=$const. Here $E(k_0)$ is the energy spectrum at some $k=k_0$ (i.e. for $q=1$) in the energy containing interval. 
 
 Now, \Eqs{final} and \eqref{Psi1}  give
\begin{align}\begin{split}\label{repB}
 \cancel{\frac{\partial \C \Psi ^2 }{\partial \tau}}+ & C (\B q) q^{8/3}\Big[\frac{11   }{2 q^2} \, (\B q\cdot \B \nabla_{\B q} ) \, \Psi ^3  -    |\B \nabla_{\B q} |^2 \, \Psi^3    \Big]   =  
 -\frac {\~\Omega   \, \Psi ^2 \, q_\|^2}{q_\|^2+ q_\times^2}   \,,\quad  \B \nabla_{\B q}\= \frac{d}{d\B q}\,, \\
 C(\B q)= & \frac{2C_1} { 2+ 3 (\B q\cdot \B \nabla_{\B q})\Psi}\,, \quad \~\Omega= \frac{\Omega}  {\sqrt{k_0^3 E(k_0)}}\,, \quad \tau=\frac t   { \sqrt{k_0^3 E(k_0)}}\,, \ q_\times= 
\frac{ \Omega\sb{ns} }{(k_0 U\sb{ns})}\,,\
\end{split}\end{align}
where we neglected the $\B q$-derivative of slow function $C (\B q)$ and took into account that in 2D $\B \nabla_{\B q}\cdot (\B q/q^2)=0$. Here, for the shortness we skip the index $j$, keeping in mind that this equation is valid for both the superfluid (with $j=$s) and  for the normal-fluid component (with $j=$n).   
 After explicit  differentiation and division of the resulting equation by $\Psi_j$ we get  
\begin{equation}\label{repC} 
  \cancel{2\,\frac{\partial \C \Psi_j }{\partial \tau}}+ 3~  C (\B q)\,q^{8/3}\Big[\frac{11 \Psi }{2 q^2} \, (\B q\cdot \B \nabla_{\B q}) \, \Psi  - \Psi  |\nabla_{\B q}|^2 \, \Psi - 2\, |\B \nabla_{\B q}  \Psi |^2 \Big]   =  
 -\frac {\~\Omega _j \, \Psi_j \, q_\|^2}{q_\|^2+ q_\times^2}   \ .
\end{equation}
 We see that the gradient of function $\Psi(\B q)$  is present in each term in the square brackets in the left-hand-side of \Eq{repC}. Therefore for zero right-hand-side (RHS), this equation admits a solution $\Psi(q)=$const.
 
 The dimensionless parameters $\~\Omega_j$ and $q_\times$ quantify the mutual friction force.
In typical laboratory experiments~\cite{WG-2017,WG-2018},  $q_\times$ belongs to the interval $q_\times \in [1,\ 8]$, while $\~\Omega_n\in [3\,, \ 12]$. In DNS\,\cite{He4-PRL-DNS,CF-DNS}, $q_\times\approx 1.3$,  $\~\Omega_n\simeq 3$. Having in mind comparison of these results with ours we will analyse \Eq{repC} in the following range of  parameters:
 \begin{equation}\label{par1}
 q_\times \in [1,\, 25]\,, \   \~\Omega_n \in [2,\, 15]\,, \  C_1 \in [0.1,\, 0.5]\, .
\end{equation} 
  For $T\approx 1.87\,$K, we approach so-called symmetric case with $\rho\sb n\approx \rho\sb s$. Furthermore we can reasonably assume that both components are equally forced, $E\sb{n}(\B k_0)=E\sb{s}(\B k_0)$. In this case we can put   $j=$s=n, considering one equation $E(\B k)=E\sb{n}(\B k)=E\sb{s}(\B k)$ instead of two equations for $E\sb{n}(\B k)$ and  $E\sb{s}(\B k)$ separately. 

    \section{\label{ss:isotrop2D}  Qualitative analysis of   anisotropic 2D energy rate equation} 
   
   The presence of the mutual friction term in the RHS of \Eq{repC} leads    to decay of function $\Psi$. As a result, $E(k)$ decays  even faster than in K41 regime $  E(q)\propto q^{-8/3}$, being  very far from the thermodynamic equilibrium with $E(k)\propto k$.   In this regime, we can use a simpler algebraic closure for the energy flux \eqref{K41A} instead of the differential closure \eqref{K41D}. This  is equivalent to   neglecting two last terms in the square brackets of \Eq{repC}. After  division of the resulting equation by  $\Psi$ we get the simplified version of the energy rate \Eq{repC}:
 \begin{equation}\label{simple}
  (\B q\cdot \B \nabla_{\B q} )\Psi (\B q )   =-  \frac {2\,\~\Omega  \,  \, q_\| ^2}{33\,C_1\, q^{2/3}
    (q_\| ^2+ q_\times^2)}\ .
 \end{equation}
 Here we took for simplicity $C(\B q)=C_1$. 
 
   For   very small $q_\|\ll q_\times$, in a zero-order approximation we can neglect the mutual friction term in the RHS 
    of \Eq{simple}. Then $\Psi(q_\|,q_\perp)\simeq \Psi(0.0)=$const.  Note, that $\Psi(q_\|,q_\perp)$ is even function of $q_\|$ and therefore has an extremum (presumably maximum) for $q_\|=0$. This allows us to hope that $\Psi(q_\|,q_\perp)$ can be roughly factorised as  $\Psi(0,q_\perp) \=\Psi_\|(0) \Psi_\perp (q_\perp)$ with  $\Psi_\|(0)=1$.   In a more extended region, say, up to $q_\|\lesssim q_\times$,  the mutual friction term becomes important and  $\Psi_\|(q_\|)$  decays fast with increasing  $q_\|$.  As we show below,  a significant (or complete) decay of $E(q_\|,q_\perp)$ takes place in a narrow, compared to $q_\perp$, range of $q_\|$. Therefore, in this case we can interpret this phenomenon as a one-dimensional problem along $q_\|$, in which $q_\perp$ and $\Psi_\perp (q_\perp)$ can be considered as parameters.  From the formal viewpoint, it means that  we can  accept (as a reasonable approximation)   a factorization 
     \begin{equation}\label{fac}
         \Psi(q_\|,q_\perp) \approx \Psi_\|(q_\|) \Psi_\perp (q_\perp)\,, 
     \end{equation} 
     neglect $q_\perp$-derivative and approximate $q$ as $q_\perp$. All these simplify \Eq{simple} as follows:   \begin{equation}\label{simple1}
 \frac{d \Psi_\| ( q_\| )}{d q_\|}   =-  \frac {2\,\~\Omega  \,  \, q_\| }{33\,C_1\, \Psi_\perp (q_\perp) q_\perp^{2/3}
    \big(q_\| ^2+ q_\times^2\big )}\ .
 \end{equation}  
To specify the boundary conditions, we introduce some $q_*$ in the beginning of the inertial interval (not necessarily equal to unity). Then, the solution of \Eq{simple1} with $ \Psi_\|(q_*)=1$     is
  \begin{equation}\label{Psi-par}
      \Psi_\|(q_\| )=1-  \frac {2\, \~\Omega  \,  \ln \big [  (q_\times^2+  q_\|^2)  \big / (q_*^2+  q_\times^2) 
     \big ] }{33\,C_1\, \Psi_\perp (q_\perp) q_\perp^{2/3}}  \ .
  \end{equation}  
 We see that both $\Psi_\|(q_\|)$ and  $\C E(\B q )\propto \Psi_\|(q_\|)$ vanish for some  $q_\| =q \sb{cr}$, for which  $ \displaystyle 2\~\Omega _j \, \ln \big [ 1+  \big (q\sb {cr} \big / q_\times\big )^2] =33 C_1 \Psi_\perp (q_\perp)q_\perp^{2/3}$ and  the RHS of \Eq{Psi-par} vanishes. This regime corresponds to so-called ``super-critical regime", first predicted in \Refn{LNV}, studied in more details in \Refn{LNS} and numerically discovered in $^3$He in \Refn{DNS-He3}.   The mutual friction affects these spectra for all $q<q\sb{cr}$ such that along the direction of the counterfow there is  no cascade-dominated $q$-range.
    
      \begin{figure} [b]
 	 \includegraphics[width=6.9  cm]{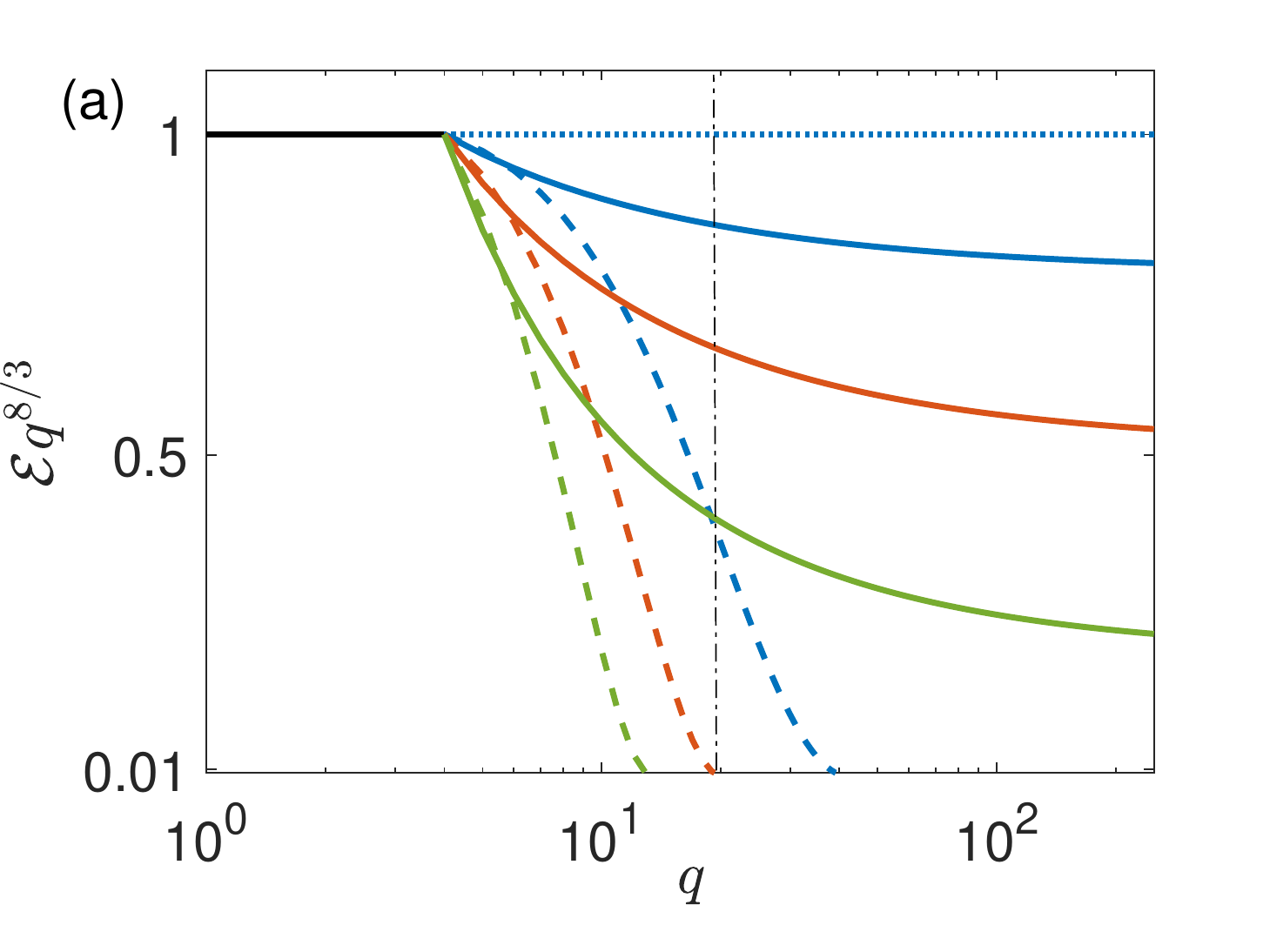}  
 		\includegraphics[width=6.9 cm]{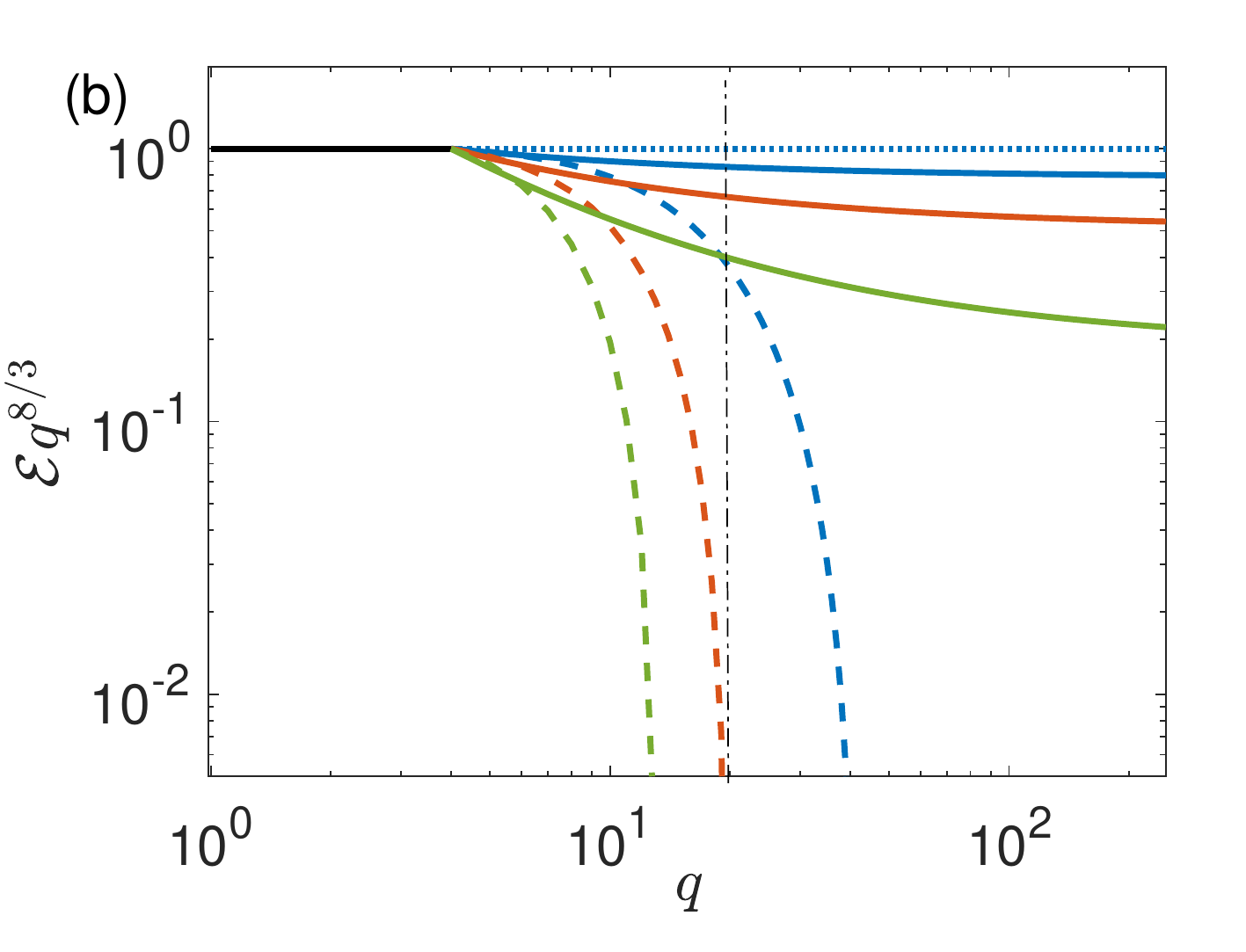} 
 \caption{ 	\label{f:1} The  K41-compensated  spectra  in  direction of the counterflow\,\Eq{Psi-par}, $q_\|^{8/3} \C E(k_\|,0)=\Psi_\|^2(q_\|)$    
  (dashed lines)  and  in the orthogonal direction, \Eq{final3B}, $q_\perp^{8/3} \C E(0,k_\perp)=\Psi_\perp^2(q_\perp)$(solid lines). The parameters of the spectra  $q_\times=20$,   $\omega\sb{dis}=0.7 \~\Omega$ and $q_*=4$. 
  Three sets of lines from top to bottom correspond  to $\~\Omega=2$ (blue lines),  $\~\Omega=5$ (red lines) and $\~\Omega=10$ (green lines).
  Note the log-linear scales in (a) and  the log-log scales in (b). Vertical black dot-dashed line denotes the $q_\times=20$.}
     \end{figure}  
    
Probably, the most straightforward way to understand the behaviour of $\Psi_\perp(q_\perp)$ is to return back to \Eq{final} and to integrate it over $k_\|$ for fixed $k_\perp$. Then, the flux term in $k_\|$ direction $\propto \partial [\dots]/ \partial k_\|$, responsible for the energy redistribution over $k_\|$ vanishes and we get the rate equation for $ ^\perp \! E(k_\perp)\= \int E(k_\|,k_\perp) d k_\|$: 
 \begin{subequations}\label{final2} 
 \begin{equation}\label{final2A} 
     \cancel{\frac{\partial \  ^\perp \!  E( k_\perp,t)}{\partial t}}  -   \frac d{d k_\perp} \int  \big \{ \dots \big \} dk_\perp  =- \omega \sb{dis} \  ^\perp \! E( k_\perp,t)\,,
 \end{equation} 
    with the same expression in $\big \{ \dots \big \}$ as in \Eq{final}. The choice of the effective  frequency $\omega\sb{dis}$, responsible for the  dissipation  by mutual friction of the energy   $ ^\perp \!  E( k_\perp,t) $ in the RHS of \Eq{final2A}, is very delicate. If we  assume that the loss of the energy  $ ^\perp \!  E( k_\perp,t)$ at some given $k_\perp$  is due to the mutual friction at the same  $k_\perp$ and all $k_\|$, then
   \begin{equation}\label{final2B} 
       \omega\sb{dis} =    \~\omega\sb{dis}\ , \quad \~\omega\sb{dis}\= \~\Omega \int   \frac{\Psi_\|^2 (q_\|)\,  q_\|^2 \, dq_\|}{q_\|^2+q_\times^2} \Big /  \int    \Psi_\|^2 (q_\|) dq_\|\ .
    \end{equation}\end{subequations}
 However, the main part of the energy  $^\perp \!  E( k_\perp,t)$ is localized in the range of relatively small $k_\|$ and the energy outflux from this region is suppressed in our model by the symmetry, because $\B \nabla_{\B k}\dots =0$   for $ \B k=\{0,k_\perp\} $ and small for small $k_\|$. It is then reasonable to assume that $0.5< \omega\sb{dis}/ \~\omega \sb{dis}<1$. In its turn, the ratio  $\~ \omega\sb{dis}/\~\Omega$  in the range of parameters \,\eqref{par1}  is close to unity.  Therefore, considering   $\omega\sb{dis}$ as a phenomenological parameter, we expect that  $ 0.5< (\omega\sb{dis}/  \~\Omega) <1$.  
 
Analysing \Eqs{final2} in the same manner as we did for \Eq{final}, we arrive at the following equations for $\Psi_\perp(q_\perp)$,
similar to \Eq{simple1} for $\Psi_\|(q_\|)$:
\begin{subequations}\label{final3} 
\begin{equation}\label{final3A} 
 \frac{d \Psi_\perp ( q_\perp )}{d q_\perp}   =-  \frac {2\,\omega\sb{dis}   }{33\,C_1\,  q_\perp^{5/3} 
   }\ .
    \end{equation} 
 
 Its solution with the boundary condition  $ \Psi_\perp(q_*)=1$  is
  \begin{equation}\label{final3B} 
      \Psi_\perp (q_\perp )=1-  \frac {4\,  \omega\sb{dis}  \,  \big  (  q_*^{-2/3}-   q_\perp^{-2/3} 
     \big ) }{99\,C_1 }  \ .
  \end{equation}
  \end{subequations}  
This equation, together with \Eqs{Psi1}, \eqref{fac} and  \eqref{Psi-par},   results  in the  semi-quantitative representation of the anisotropic 2D energy spectrum of the unbounded  counterflow turbulence with the  axial symmetry: 
  \begin{equation}\label{RES} 
  E(q_\|,q_\perp)\simeq \frac {E(q_*)} {q^{8/3}}\big[ 1-  \frac {2\, \~\Omega  \,  \ln \big [ (q_*^2+  q_\|^2) \big / (q_\times^2+  q_\|^2)\big ] }   {33\,C_1\, \Psi_\perp (q_\perp) q_\perp^{2/3}}\Big]^2 
  \Big[1-  \frac {4\,  \omega\sb{dis}  \,  \big  (  q_*^{-2/3}-   q_\perp^{-2/3} 
     \big ) }{99\,C_1 } \Big]^2 \ .
     \end{equation}

  The explicit form \eqref{RES} for the anisotropic energy spectra of counterflow turbulence is the main analytical result of our paper.

       To explore   the  form of the 2D-energy spectrum\,\eqref{RES},  we plot in \Fig{f:1} 
           the cross-sections of the K41-compensated  spectra  in  direction of the counterflow, $k_\|^{8/3} \C E(k_\|,0)=\Psi_\|^2(q_\|)$ 
  [\Eq{Psi-par}, dashed lines]  and  in the orthogonal direction $k_\perp^{8/3} \C E(0,k_\perp)=\Psi_\perp^2(q_\perp)$  
  [\Eq{final3B},   (solid lines)].
  The log-linear scales in  \Fig{f:1}(a)  expose the details of  $k_\perp^{8/3} \C E(0,k_\perp)$, while   the log-logs scale in \Fig{f:1}(b) emphasize the strongly suppressed $k_\|^{8/3} \C E(k_\|,0)$. We see that the spectra in the counterflow direction experience fast decay and sharp cut-off, corresponding to the super-critical regime, described above. On the other hand, the spectra in the orthogonal direction decay much slower, corresponding to the so-called “sub-critical regime" \cite{DNS-He3,LNV,LNS}, which consists of two K41 scaling laws: in the range of small $q$ it has  the energy flux $\ve_0$  equal to the rate of the energy pumping, while for large $q$  it has  smaller energy flux $\ve_\infty<\ve_0$. The difference  $\ve_0-\ve_\infty$, is dissipated on the way to large $q$ due to
mutual friction. At larger $q$, the dissipation by mutual friction  is no longer efficient because
scale-independent large-$q$ asymptotic of the mutual friction frequency  $\~\Omega$ becomes finally
smaller than the K41 energy transfer frequency $\gamma(q)\simeq \ve_\infty^{2/3} q^{2/3}$. Similar effect of vanishing of the mutual friction effect at small scales was originally observed in an isotropic system in \cite{LNV}.
    
We conclude that from the viewpoint of the  qualitative analysis of the energy rate \Eq{repC}, the energy spectrum of counterflow turbulence  has a pancake form around the  counterflow direction $q_\|$. It is strongly confined in $q_\|$ direction due to the special anisotropic form of the mutual friction force, effective only for $k_\|\ne 0$. In the next section we consider the numerical solution of the model \Eq{repC} and compare the results with the qualitative predictions. 
 
     \begin{figure} [t]
      \begin{tabular}{cc  }	
  	\includegraphics[width=7cm]{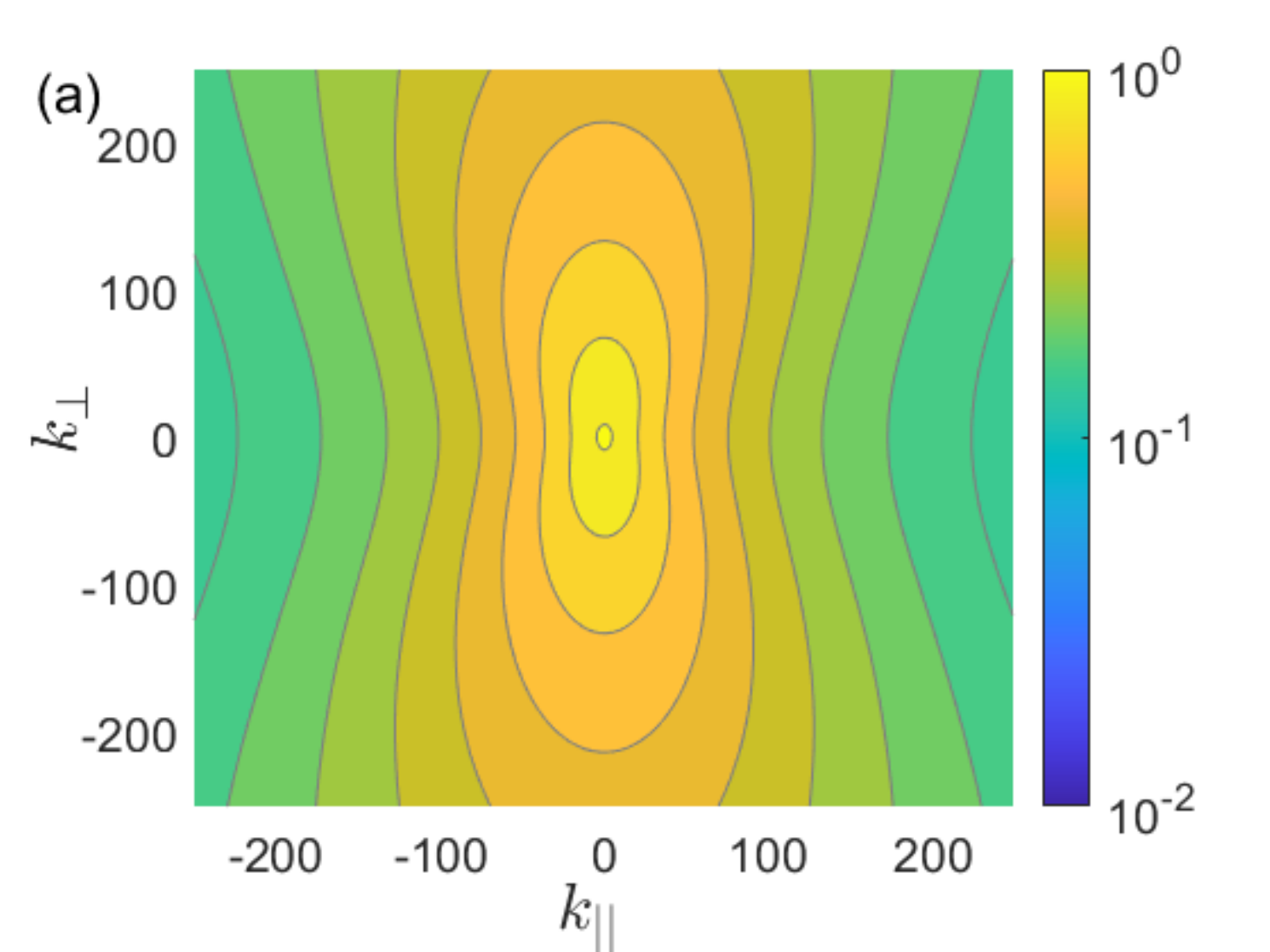} &\hskip - .6  cm
	\includegraphics[width=7cm]{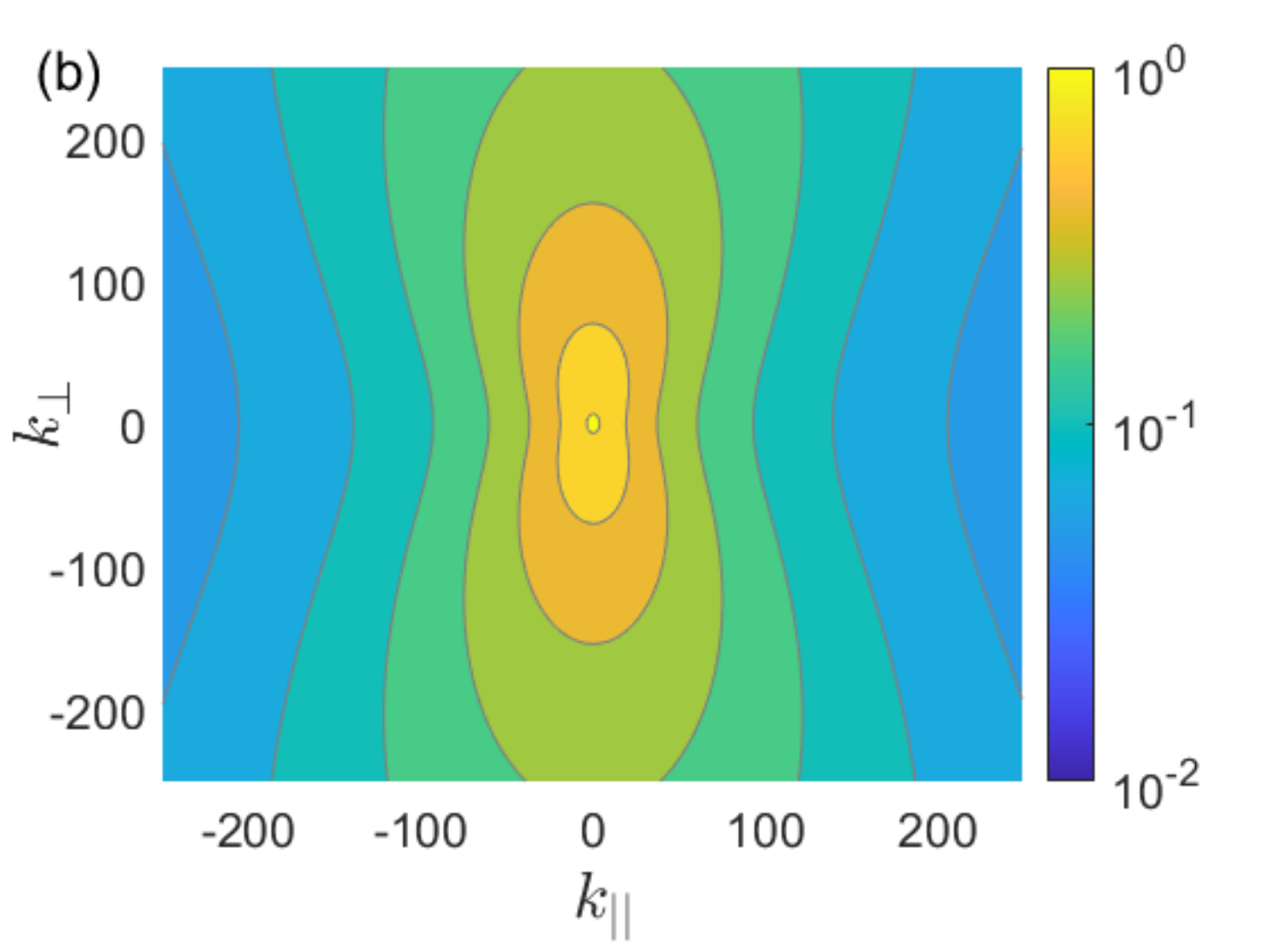}
	\\
  	\includegraphics[width=7cm]{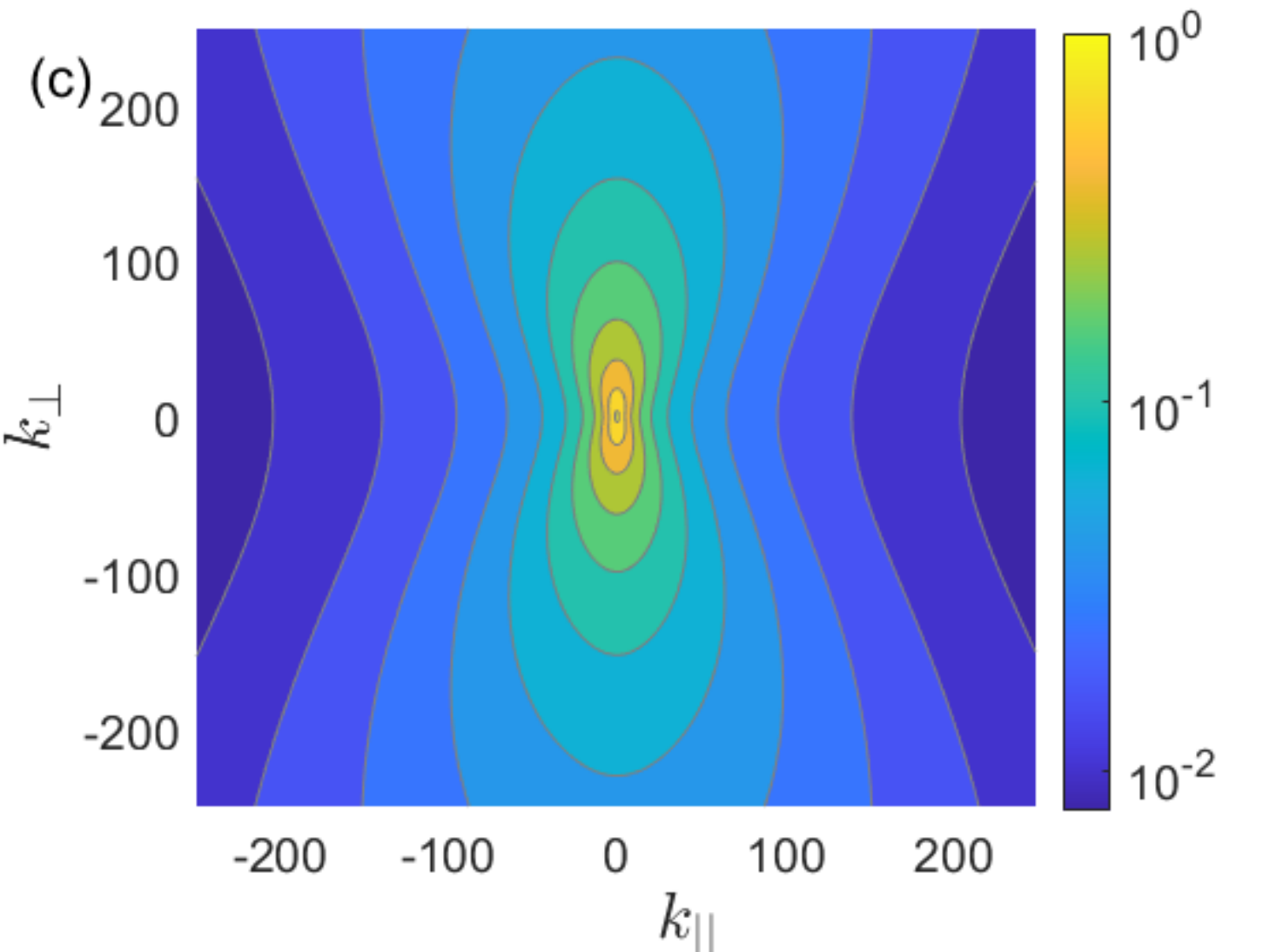}& \hskip - .6 cm
	\includegraphics[width=7.1cm]{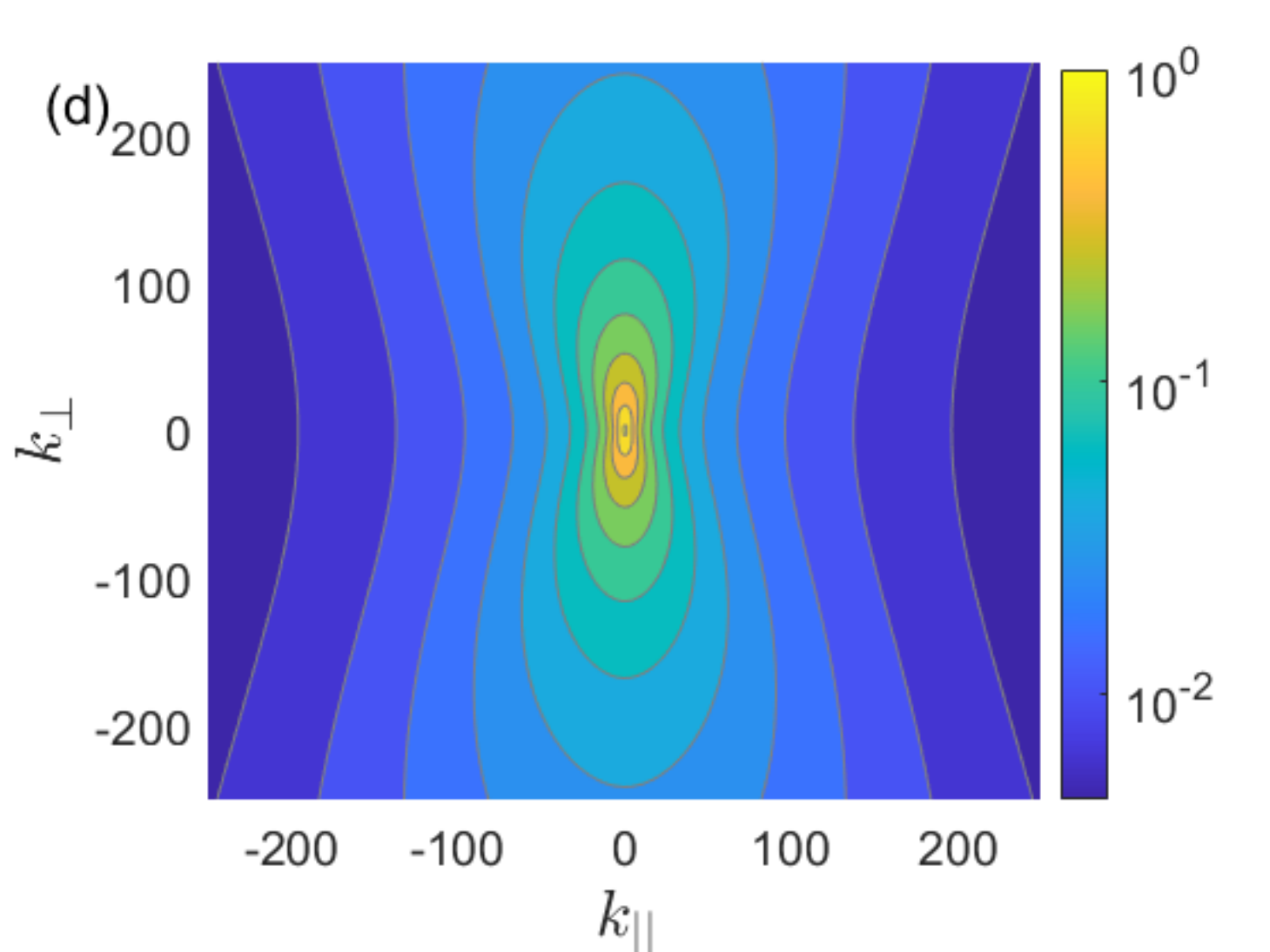} 
		\end{tabular}   
 	\caption{ 	\label{f:2}The K41-compensated 2D energy spectra $k^{8/3} \C E(\B k)$. Panels
   (a), (b): the spectra are calculated for $k_\times = 100$ and $\~\Omega=2,\ 5$ respectively. Panels (c), (d): the spectra are calculated for	$k_\times = 20$ and the same values of $\tilde \Omega$. Note logarithmic scale of the color-bars. The  contour levels are spaced  by 0.1 in (a) and (b) and by 0.2 in (c) and (d).
 	}
 	\end{figure}
                  
            \begin{figure} [t]
  \begin{tabular}{cc  }	
 \includegraphics[width=6.7 cm]{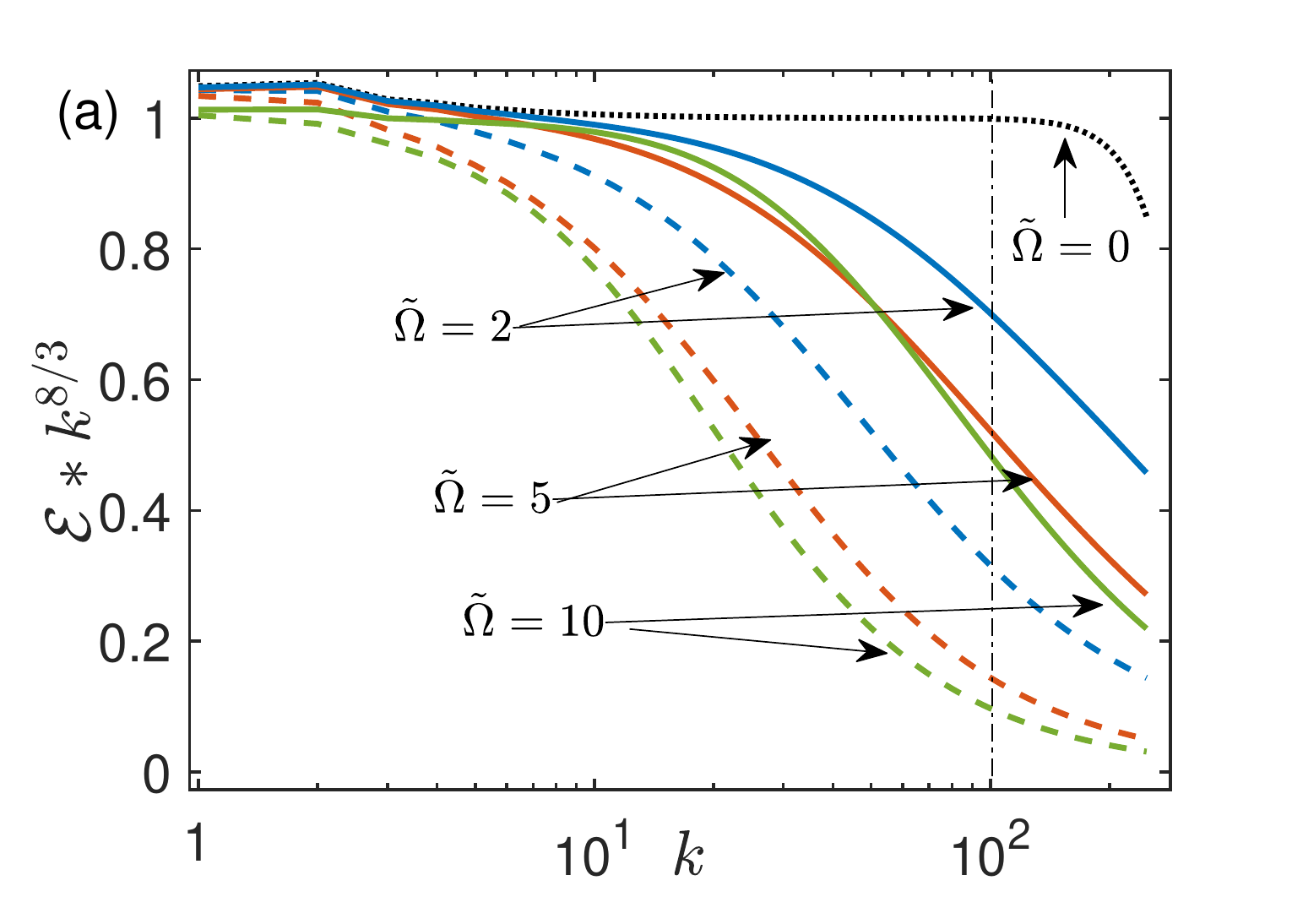}  &
 \includegraphics[width=6.7 cm]{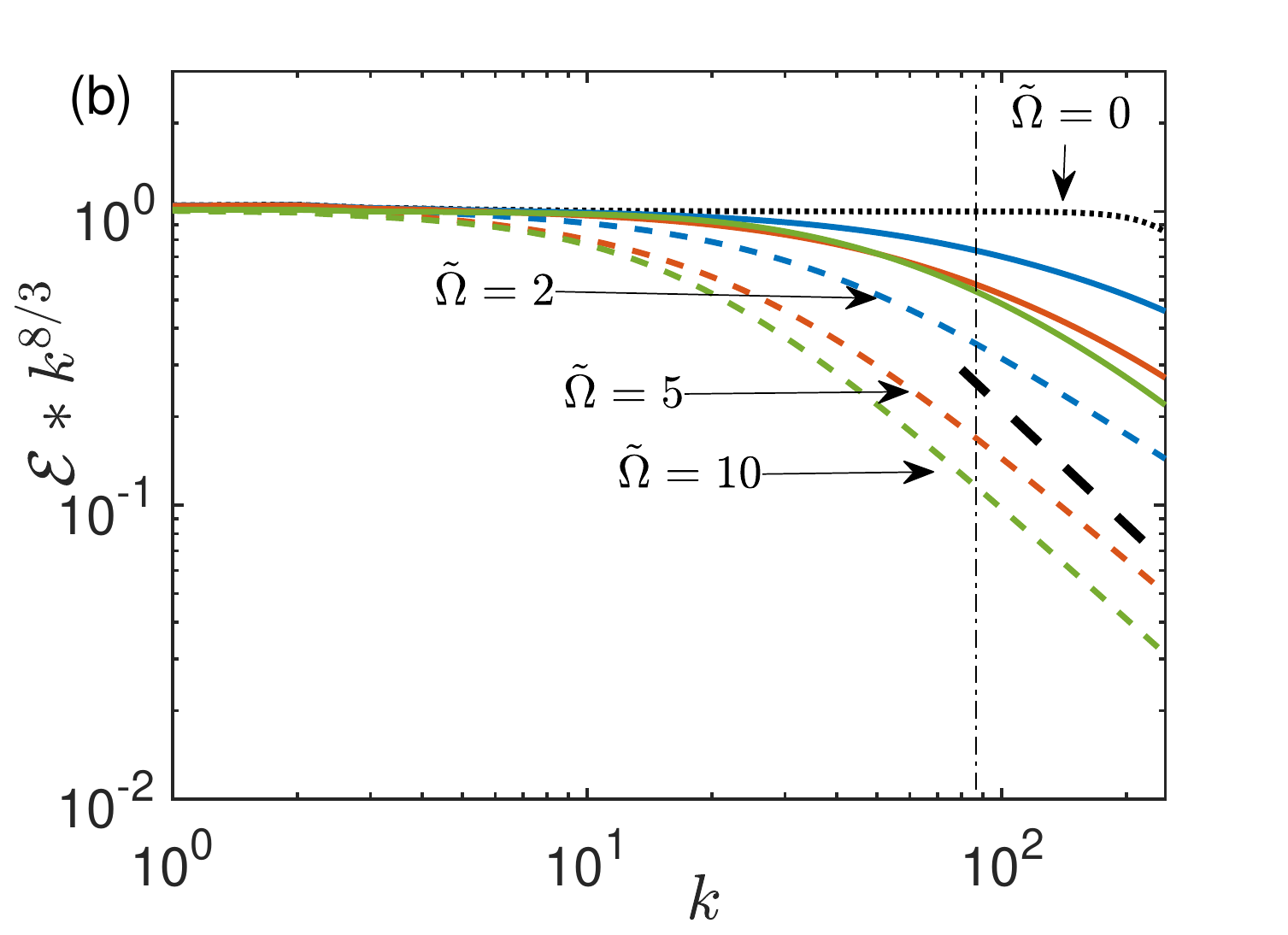}  \\ 
 \includegraphics[width=6.7 cm]{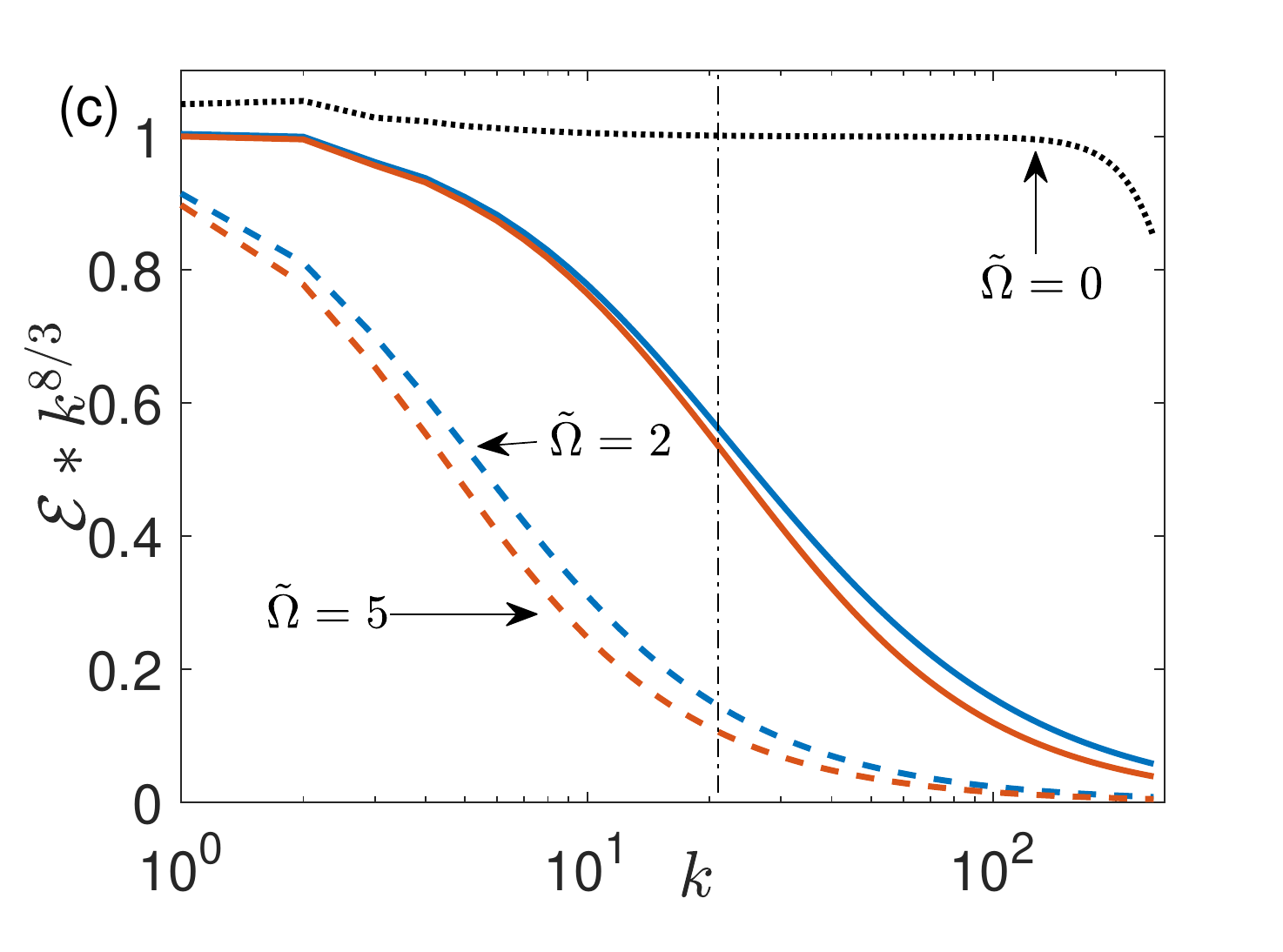}  &
\includegraphics[width=6.7 cm]{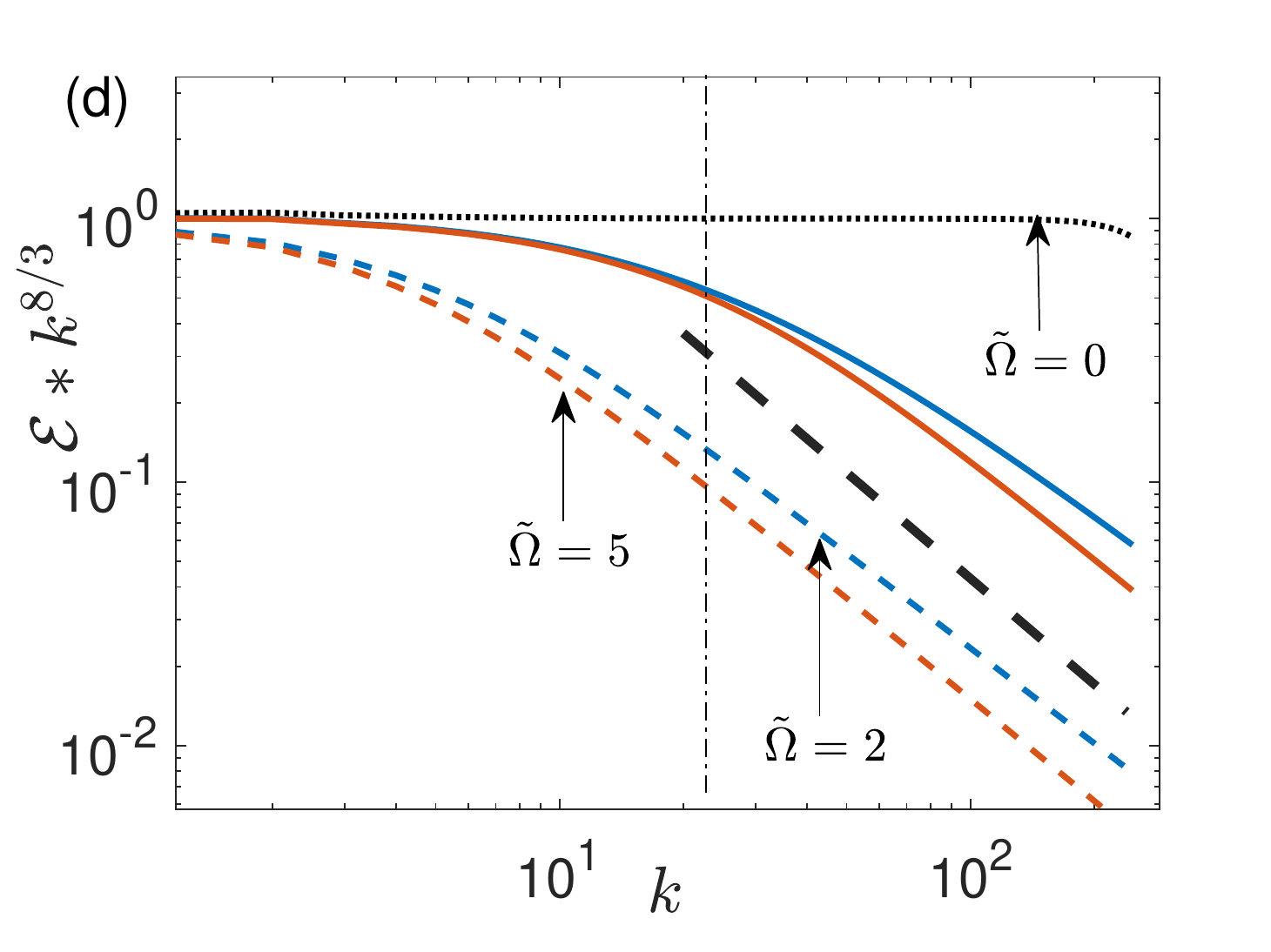} 
 		\end{tabular}   
 \caption{ 	\label{f:3} Numerical solution of \Eq{repC}.
  The K41-compensated spectra along $k_\|^{8/3} \C E(k_\|,0)$ and normal to the counterflow direction $k_\perp^{8/3} \C E(0,k_\perp)$. 
    The values of $\tilde \Omega$ are indicated in the figure. In (a) and (b) $k_\times=100$, in (c) and (d) $k_\times=20$.  The reference case  $\~\Omega=0$ (no mutual friction) is plotted in all panels by a black dotted line.   Vertical dot-dashed lines denote the position of the crossover wavenumber $k_\times$.  Black thick dashed lines  in (b) and (d)  denote ${\C E} \propto k^{-4}$ and serve to guide the eye only. Vertical dot-dashed lines indicate the position of $k_{
    \times}$.  Note the log-linear scales in (a) and (b) and the log-log scales in (c) and (d). }
      \end{figure}

      \section{\label{s:sol} Numerical  solution  of  energy rate equation and  discussion  } 
 The equation  \eqref{repC} (with the replacement $\B q\to \B k$) was solved numerically   as a time evolution on the $500^2$-grid with the self-consistent form of $C_1(\B k)$ given by \Eq{repB}.
 We used the initial condition $\Psi(\B k,0)=1$ for all $\B k$.   To reach the stationary solution, we added a forcing term with small amplitude $f_0=0.005$, acting in first four modes  $k=\sqrt{k_\|^2+ k_\perp^2}\le k_*=4$ and an artificial exponential dumping term, acting at the edges of the grid. After a short transient period, a steady-state solution for $\Psi(k_{\|},k_{\perp})$ was obtained. We have verified that this solution is insensitive to the details of forcing and artificial dumping, as long as the stationary solution is reached.  

 The contour plots of the 2D energy spectra for several sets of  parameters of the problems, $\~\Omega$ and $k_\times$, are shown \Fig{f:2}. The spectra are clearly confined along $k_\|$, more strongly with increasing $\tilde \Omega$ and decreasing $k_{\times}$. Indeed, according to \Eq{repB}, larger $\~\Omega$ enhances the  mutual friction, while smaller $q_\times$ increases the range in $k$-space where the mutual friction is important.  

The cross-sections of the 2D  compensated energy spectrum $k^{8/3} \C E(\B k)=|\Psi(\B k)|^2$ are shown in \Figs{f:3} (a) and (b) for $k_\times=100$ and in \Figs{f:3} (c) and (d)  for $k_\times=20$. The spectra $\C E(k_\|,0)$ along $k_\|$, are shown by dashed  lines and $\C E(0, k_{\perp})$ along $k_{\perp}$,  by solid lines. Similar to \Fig{f:1}, we plot the  spectra both in the log-linear scales to emphasize the details of the orthogonal spectra, and in the more conventional log-log scales.

 We see that spectra along the counterflow direction experience fast decay, while  the energy cross-sections in the orthogonal direction decay much slower. For $k_{\times}=100$, the orthogonal spectra have some interval of the cascade-dominated range with near-K41 scaling that is shorter for larger $\tilde \Omega$.  The spectra along $k_\|$ do not have such an interval for these parameters. For $k> k_{\times}$, all spectra have similar power-law  behavior, which we discuss below.  For $k_{\times}=20$, the spectra quickly saturate with increasing $\tilde \Omega$ and are almost completely in the mutual-friction-dominated range. However, due to self-consistent closure for the energy flux, the spectra  do not become super-critical, as  in the analytic solution.  

\begin{figure} [t]
 	 \includegraphics[width=6.8  cm]{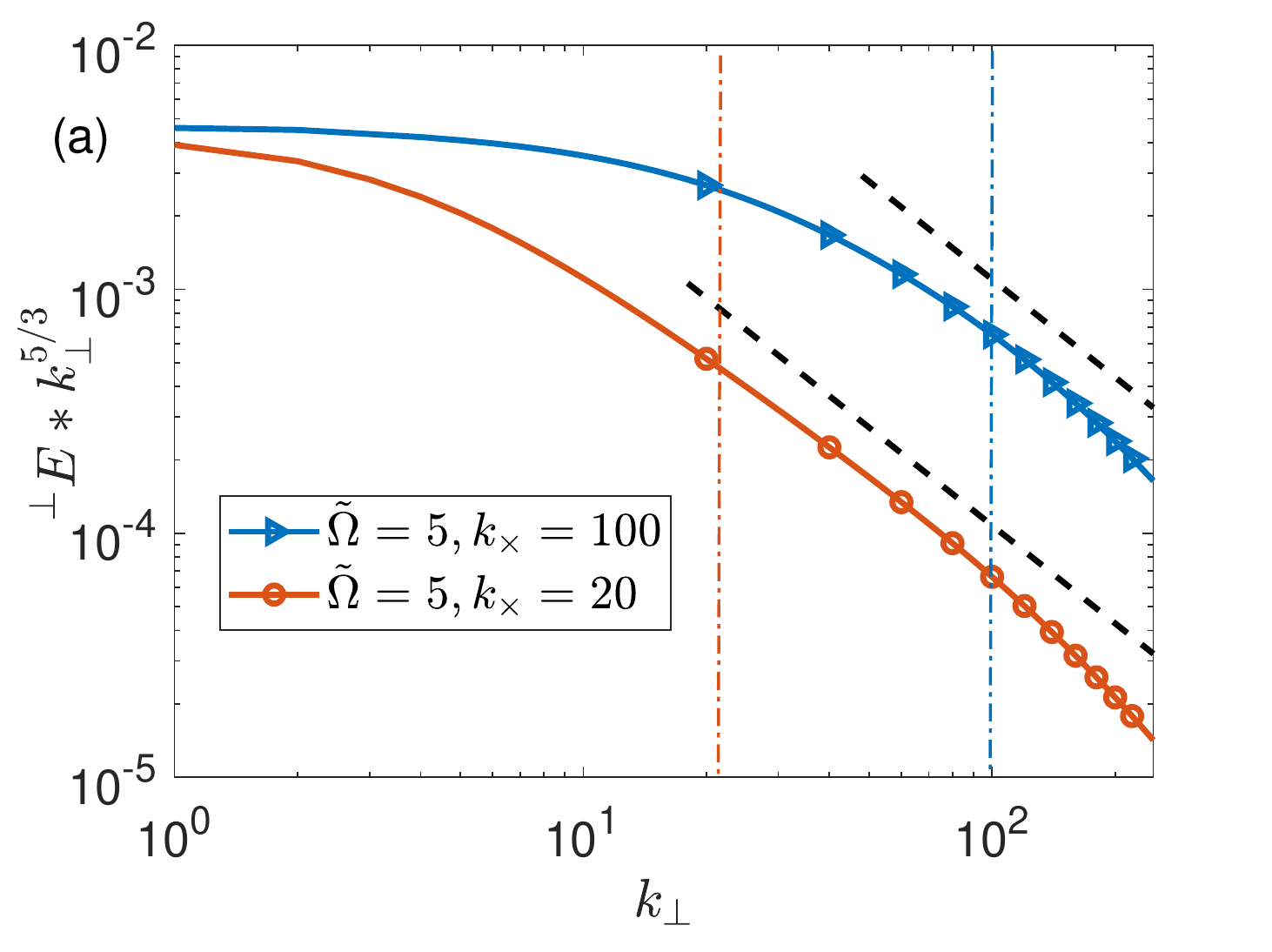}  
 		\includegraphics[width=6.9 cm]{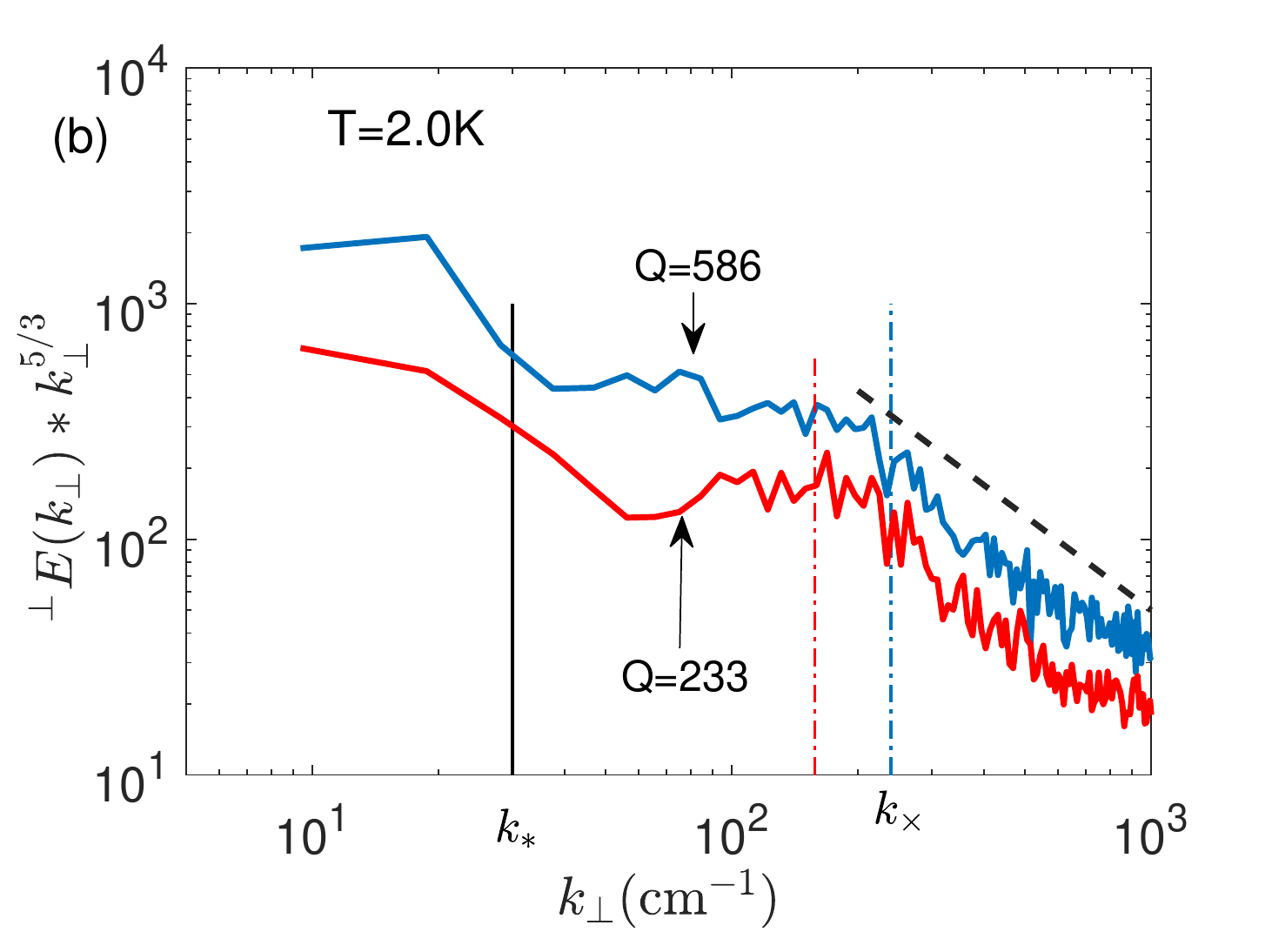} 
 \caption{ 	\label{F:4} Comparison  of the  theoretical  and experimental~ K41-compensated   1D plane-averaged energy spectra $^\perp\!E(k_\perp) ~k_{\bot}^{5/3}$. Panel (a): the theoretical spectra [ \Eq{comp}],  for $\tilde \Omega=5$ and two cross-over wavenumbers. Panel (b): experimental spectra measured by molecular-racer velocimetry \cite{WG-2018}, at $T=2.0$ K and two heat fluxes. The vertical dot-dashed lines of matching colors in both panels denote the position of the corresponding $k_{\times}$.  Black dashed lines denote critical scaling  $^\perp\!E(k_\perp)\propto k_\perp^{-3}$.}  
     \end{figure}

 Another result of principle importance is the universality of the  scaling exponent $x\sb{cr}=4$ of both longitudinal and transverse cross-sections  of the energy spectra, $\C E(k_\|,0)\propto k_\|^{-x\sb{cr}}$,  $\C E(0,k_\perp)\propto k_\perp^{-x\sb{cr}}$ 
shown in \Figs{f:3}(b) and (d) by thick black dashed lines.  The exponent $x\sb{cr}=4$ in 2D energy spectra manifests itself in the so-called critical energy spectra, appearing in the  regimes with strong enough mutual friction. The critical energy spectrum
separates the sub-critical and the super-critical energy spectra with local and non-local energy transfer over scales\,\cite{DNS-He3}, respectively. In the critical regime, the fraction of the energy loss due to mutual friction at each scale is about the fraction of the energy transferred down to smaller scales. We argue that the critical regime is reached for $k>k_{\times}$ in the wide range of the flow parameters. This conclusion is supported experimentally: in \Refn{WG-2018}
the critical regime was observed in $^4$He counterflow for $T=1.65,\, 1.85,\, 2.00\,$K and $T=2.10\,$K. In this paper, the normal-fluid component of the counterflow is probed by He$_2^*$ molecular tracer-line tracking technique, allowing to measure 1D plane-averaged energy spectrum $^\perp\!E(k_\perp)$, connected to studied here 2D-spectra $E(k_\|, k_\perp)$ as follows
\begin{equation}\label{comp}
^\perp\!E(k_\perp)=\int E(k_\|, k_\perp) dk_\|\ .
\end{equation}
To compare our theory and experiment\,\cite{WG-2018}, we plotted in \Fig{F:4}(a) the K41-compensated spectra  $k^{5/3}_\perp\,^\perp\!E\sb{th}(k_\perp)$,  for $\~\Omega=5$ and two different   $k_\times$.   In \Figs{F:4}(b) we plotted the experimental spectra $k^{5/3}_\perp\,^\perp\!E\sb{exp}(k_\perp)$, measured  for $T=2.00\,$K and two heat fluxes. In both theoretical and experimental spectra, we clearly see two regimes with different apparent scalings: i) in the region of small $k_\perp$ (roughly below and about $k_\times$) -- non-universal apparent exponents, that depend on the flow parameters and are close  to the K41 scaling (almost horizontal lines for K41 compensated spectra) and ii) universal scaling with exponents, close to the critical value $\~x\sb{cr}=3$ for $k_\perp>k_\times$. Note, that 1D exponents differ by unity from their 2D counterparts, e.g.  in 1D, the K41 scaling exponent $\~y\Sb{K41}=5/3$ and $\~x\sb{cr}=3$, while 
 in 2D,  $y\Sb{K41}=8/3$ and $x\sb{cr}=4$.  We, therefore, infer that  our theory reproduces two scaling ranges, previously found in laboratory experiments\,\cite{WG-2018}: the cascade-dominated range in the range of small $k$ with scaling $^\perp\!E(k_\perp)\propto k_\perp^{-y}$, close to the K41 exponents $ y \gtrsim  \dfrac 53$ and the mutual-friction dominated range with the critical scaling $^\perp\!E(k_\perp)\propto k_\perp^{-3}$.
 
 	     \section{\label{s:sum} Summary}
 	       We developed an analytic theory of  energy spectra  in the thermally-driven  turbulent counterflow of superfluid $^4$He, which generalises the L'vov-Pomyalov theory of counterflow turbulence\,\cite{LP-2018} to strongly anisotropic case.  The  theory is based on the gradually-damped\,\cite{CF-DNS}
	coarse-grained  \Eqs{NSE} of the incompressible superfluid turbulence\,\cite{He4,DNS-He4,DNS-He3}
and the novel anisotropic, self-consistent differential closure\,\eqref{closure}  for the vector of the turbulent  energy flux $\B \ve (\B k)$. This closure combines the Kolmogorov-1941 dimensional reasoning\,\cite{Fri}, the Leigth-1968 differential form\,\cite{Leith} to account for possibility of the thermodynamic equilibrium and L'vov-Pomyalov-2018  self-consistent closure for the energy flux\,\cite{LP-2018} that   accounts for the dependence of the energy flux on the local slope of the energy spectrum in the window of its locality.   In addition, the  suggested closure prescribes the orientation  of the vector of the energy flux $\B \ve (\B k)$ in the steepest-decent direction of 3D turbulent energy spectra $F(\B k)$ toward its thermodynamic equilibrium: $\B \ve (\B k)\| \B \nabla_{\B k} F(\B k)$. 
	
Similar to previous theories\cite{decoupling,LP-2018}, the  important element  of our theory is the anisotropic cross-correlation function\,\eqref{cross} between the  superfluid and normal-fluid velocity components. This function determines the rate of energy dissipation by the mutual friction in the final energy rate equation \eqref{final}. 

	Detailed   analysis of \Eq{final} leads to the analytic \Eq{RES} for the energy spectrum that  describes its strong  suppression  
	 with respect to the classical fluid counterpart. The spectra are non-scale-invariant, and strongly depend  on the temperature and the counterflow velocity in the wide range of these  parameters.  The resulting energy spectra of the normal-fluid and superfluid components are  strongly confined in the direction of the  counterflow velocity.  This conclusion is  supported by the  numerical solution of the  energy-rate \Eq{final} and by the direct numerical simulation of the coarse-grained \Eqs{NSE} for the counterflow turbulence 
	 \cite{He4-PRL-DNS,CF-DNS}. Our theory explains the critical scaling behaviour with the exponent $\~x\sb{cr}=3$ at $k>k_{\times}$, found in the experiment\,\cite{WG-2018}, that is  insensitive to the flow parameters.

	We, therefore, hope that the suggested theory captures the basic physics of
the counterflow turbulence and describes the dependence of the anisotropic energy spectra on the main flow parameters.

\vskip6pt
\enlargethispage{20pt}

\dataccess{This article has no additional data.}

\aucontribute{
All authors contributed equally.
}

\competing{ The authors declare that they have no competing interests.}

\funding{ YL gratefully acknowledges support from  Office of Naval Research (grant N00014-17-1-2852) and National Science Foundation, Division of Mathematical Sciences (DMS) (award 2009418).
SN was supported by Simons  Foundation Collaboration grant Wave Turbulence (award ID 651471).
}



\end{document}